\documentclass[a4paper,11pt]{article}
\usepackage{latexsym}
\usepackage{cite}
\usepackage{epsfig}
\usepackage{amssymb}
\usepackage{latexsym}
\usepackage{epsfig, subfigure, subfloat, graphicx,float}
\usepackage{mathrsfs}
\usepackage{amsmath}
\usepackage{mathtools}
\newcommand{\abs}[1]{\left\lvert#1\right\rvert}

\newcommand{\f}[2]{\frac{#1}{#2}}
\newcommand{\mk}[1]{\left( #1 \right)}
\newcommand{\kk}[1]{\left[ #1 \right]}
\newcommand{\ck}[1]{\left\{ #1 \right\}}
\newcommand{\be}{\begin{equation}}
\newcommand{\ee}{\end{equation}}
\newcommand{\Mpl}{M_{\rm P}}
\newcommand{\e}{\epsilon}
\newcommand{\fact}[1]{#1\mathpunct{}!}
\author{H. Mohseni Sadjadi\footnote{mohsenisad@ut.ac.ir}  and V. Anari\footnote{v.anari@alumni.ut.ac.ir}
\\ {\small Department of Physics, University of Tehran,}
\\ {\small P. O. B. 14395-547, Tehran 14399-55961, Iran}}
\title{End of the constant-roll inflation, and the reheating temperature}
\begin{document}
\maketitle
\begin{abstract}
By extending the potential, we propose a mechanism for the end of the constant-roll inflation and the subsequent reheating phase in the FLRW space-time.
Based on astrophysical data, we estimate the Universe reheating temperature.
\end{abstract}

\section{Introduction}
To solve cosmological problems like the horizon and flatness problems and the absence of magnetic monopoles, the inflation model has been proposed \cite{inf1}. This theory asserts that the early Universe has undergone a period of accelerated expansion\cite{inf2,inf3,inf4,inf5,inf6,inf7}. Responsible for this positive acceleration may be a scalar field (dubbed inflaton) whose quantum fluctuations were the seeds of the structures formation. Depending on the inflaton potential and its interactions (e.g. gravitationally non-minimal interactions) various inflationary models have been considered in the literature. One of the first models is the slow-roll where the scalar field slowly rolls down a concave potential during the inflation and eventually oscillates around its minimum.
A slowly rolling scalar field, in addition to being able to provide enough e-folds to solve the cosmological problems, determines the scalar and tensor perturbations in agreement with recent observations. When the slow-roll approximation fails, the inflation ends and finally, the scalar field decays to relativistic particles during the reheating \cite{reh1,reh2,reh3} (or preheating \cite{pre1,pre2,pre3}) era. In the warm inflation context, inflation and relativistic particles production occur in the same era \cite{warm1,warm2,warm3,warm4,warm5}.

Recently a new inflationary model, in which the time evolution of the scalar field ($\phi$) is governed by the equation $\ddot{\phi}=\beta H \dot{\phi}$ was introduced in \cite{cr1}. $H$ is the Hubble parameter and a dot denotes time derivative in the Friedmann-Lema$\hat{\i}$tre-Robertson-Walker (FLRW) space-time.  For $\beta\ll 1$ the slow-roll approximation is recovered and for $\beta=-3$, the ultra-slow-roll limit is obtained. $\beta+3\neq 0$ shows deviation from a completely flat potential. The constant-roll has been widely studied in the literature \cite{cr2,cr3,cr4,cr5,cr6,cr7,cr8,cr9,cr10,cr11,cr12}. This model may be employed to describe non-Gaussianity generation, and also the growing of curvature perturbations in super-horizon scales \cite{cr1}, and also the production of primordial black holes which are a candidate for dark matter\cite{b1,b2}.

Note that the inflation and reheating are related together via the scalar field which is responsible for both of them. The scalar potential must be specified such that it explains the inflation as well as the inflation exit, and also the subsequent reheating period. The initial conditions of the reheating era are derived from the end of inflation, and therefore the reheating temperature depends on the model used for inflation. The number of e-folds, between when our visible Universe exited the horizon and the end of inflation, restricts the inflationary parameters. This e-folds number is also related to the effective equation of state (EoS) parameter in the reheating era, and to the reheating temperature which put some more constraints on the inflation model \cite{ref1,ref2,ref3,ref4}. This approach has been used in \cite{ref1}, to study constraints to the parameter space for natural inflation and Higgs-like inflation models. The dependence of the reheating temperature on the inflation model parameters was also discussed in \cite{sad2} for a nonminimal derivative coupling model, and in \cite{sad3} for a modified teleparallel inflationary model.

In the slow-roll model, by the evolution of the inlfaton, the conditions required for the slow-roll cease and the inflation ends. The inflaton oscillates coherently around the minimum of its potential and reheats the Universe via preheating and reheating processes. To consider the reheating era, one must extend the model for example by considering inflaton decay during inflation (like warm inflation \cite{warm1,warm2,warm3,warm4,warm5}), or assuming a more general potential such that inflation is initially in line with the constant-roll potential, but by the evolution of the field, the potential changes and allows the inflation exit and occurrence of an oscillatory phase. We adopt the latter, and by modifying the potential, study the exit of inflation and also the reheating of the Universe. To obtain the reheating temperature, by investigating the Universe evolution from the horizon exit of a pivot scale until now, and astrophysical data, we estimate the reheating temperature in the constant-roll approach.

\section{Constant-roll inflation}
We consider the action
\be\label{1}
S=\int d^4x \sqrt{-g} \kk{\f{\Mpl^2}{2}R -\f{1}{2}g^{\mu\nu}
\partial_\mu\phi\partial_\nu \phi-V(\phi)},
\ee
in the FLRW space-time. The reduced Planck mass is $\Mpl= (8\pi G)^{-1/2}$ and $\phi$ denotes the scalar field.  Variation of the action with respect to the metric and the scalar field gives the equations of motion
\begin{align}
\label{2}H^2 &=\f{1}{3\Mpl^2}\left(\f{\dot \phi^2}{2}+V\right), \\
\label{3} \dot H&=-\f{1}{2\Mpl^2}\dot \phi^2,\\
\label{4} \ddot\phi&+3H\dot\phi+\f{\partial V}{\partial \phi}=0
\end{align}
The Hubble parameter is given by $\displaystyle H\equiv \f{\dot a}{a}$ in which $a$ is the scale factor. The slow-roll parameters are defined by
\be \label{5}
\e_1\equiv -\f{\dot H}{H^2},\quad \e_{n+1}\equiv \f{\dot \e_n}{H\e_n}.
\ee
In the slow-roll, $\epsilon_n\ll 1$ and $\ddot\phi$ in (\ref{4}), and ${\dot \phi^2}$ in (\ref{2}) are negligible. In the constant-roll we do not ignore $\ddot{\phi}$ and we generally have
\be \label{6}
\ddot\phi=\beta H\dot \phi,
\ee
where $\beta$ is a constant. For $\beta\ll 1$ the slow-roll evolution is recovered. For $\beta=3$, from (\ref{4}) we find  a completely flat potential, $\f{\partial V}{\partial \phi}=0$, corresponding to the ultra slow-roll. By employing the constraint (\ref{6}), one can analytically solve the equations (\ref{2})-(\ref{4}), and obtain the form of the potential. To do so we assume that $H=H(\phi)$. For single valued $t(\phi)$, we can write  $\displaystyle \dot H=\dot \phi \f{dH}{d\phi}$, therefore from (\ref{3}) we find
\be \label{7}
\dot \phi = -2\Mpl^2\f{dH}{d\phi}.
\ee
From (\ref{6}) and (\ref{7}) we obtain
\be \label{8}
\f{d^2H}{d\phi^2}+\f{\beta}{2\Mpl^2}H=0,
\ee
which gives the general solution of $H$ in terms of $\phi$ as
\be \label{9}
H(\phi)=C_1\exp\mk{ \sqrt{-\f{\beta}{2}}\f{\phi}{\Mpl} }+
C_2\exp\mk{ -{\sqrt{-\f{\beta}{2}}\f{\phi}{\Mpl}} }.
\ee
Now (2), (7), and (9) specify the potential:
\begin{align} \label{10}
&V(\phi)=\Mpl^2\kk{3H^2-2\Mpl^2\mk{\f{dH}{d\phi}}^2} \notag\\
&=\Mpl^2\Big[(3+\beta)\ck{C_1^2\exp\mk{\sqrt{-2\beta}\f{\phi}{\Mpl}}+ C_2^2\exp\mk{-\sqrt{-2\beta}\f{\phi}{\Mpl}}}\notag\\
&+2(3-\beta)C_1C_2\Big].
\end{align}
By inserting (\ref{9}) in (\ref{7}), we obtain an evolution equation for $\phi$ in terms of $t$, which specifies $\phi(t)$ and thereupon $H(t)$.

Particular solutions, are obtained by specifying the parameters $C_1$ and $C_2$ and $\beta$. For example by setting one of the $C_i$  to be zero, or $C_1=\pm C_2$, we get
\begin{align}
\label{11} H&=M e^{\pm \sqrt{-\f{\beta}{2}}\f{\phi}{\Mpl}},\\
\label{12} H&=M \cosh \mk{\sqrt{-\f{\beta}{2}}\f{\phi}{\Mpl}}, \\
\label{13} H&=M \sinh \mk{\sqrt{-\f{\beta}{2}}\f{\phi}{\Mpl}},
\end{align}
where $M$ is a constant. In the following we restrict ourselves to (\ref{12}), because as justified in \cite{cr1},  only (\ref{12}) may describe the inflationary era in agreement with astrophysical data.

\begin{figure}[H]
\centering
\includegraphics[width=8cm]{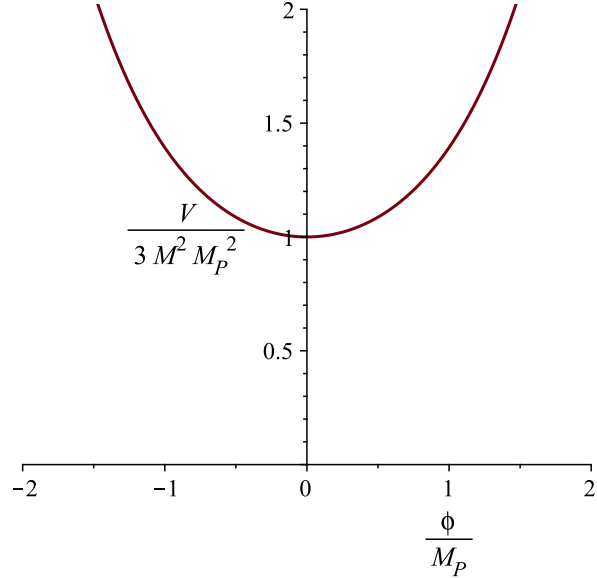}
\caption{ Potential \eqref{14} for  $\beta=-1$.}
\label{fig1}
\end{figure}

For $\beta < 0$, (\ref{12}) leads to
\begin{align} \label{14}
V(\phi)&=3M^2\Mpl^2\kk{ 1-\f{3+\beta}{6}
\ck{1-\cosh\mk{\sqrt{-2\beta}\f{\phi}{\Mpl}}} },\\
\label{15} \phi&=\Mpl \sqrt{-\f{2}{\beta}}
\ln\kk{\coth\mk{-\f{\beta}{2}Mt}},\\
H&= M \coth \mk{-\beta M t}, \\
\label{16} a &\propto \sinh^{-1/\beta} \mk{-\beta Mt}.
\end{align}
By defining a dimensionless parameter  $\tilde{\phi}=\f{\phi}{M_P}$ and expanding the potential around  $\tilde{\phi}=0$ we have :
\be\label{18}
\f{V(\phi)}{3M^2\Mpl^2} =1-\f{\beta}{6}\mk{3+\beta} \tilde{\phi}^2 + \f{\beta^2}{36}\mk{3+\beta} \tilde{\phi}^4 + \mathcal{O}\mk{\tilde{\phi}^6}
\ee
 which for $-3<\beta<0$ the potential has a minimum at $\tilde{\phi} =0$. For example for  $\beta = -1$, we have plotted \eqref{14} in terms of  $\phi$  in fig.(\ref{fig1})
But according to (\ref{18}) and also as asserted in \cite{cr1}, this solution is not an attractor for $\beta<-3$.  

\begin{figure}[H]
\centering
\includegraphics[width=8cm]{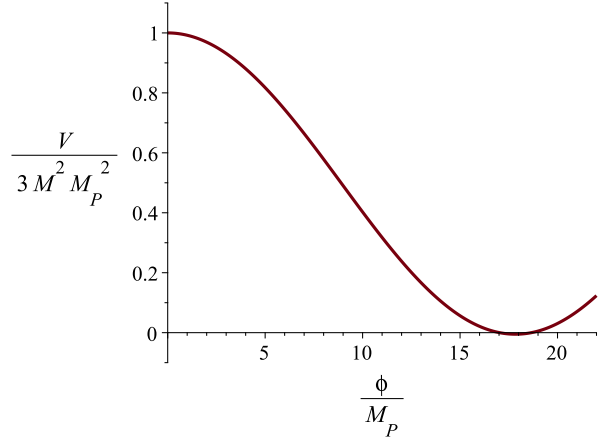}
\caption{The plot of \eqref{14} for $\beta=0.015$.}
\label{fig2}
\end{figure}

Observationally viable model which is in agreement with astrophysical data corresponds to $\beta\gtrsim 0$ (see the conclusion of \cite{cr1}). For $\beta > 0$  we have \cite{cr2}:
\begin{align}
\label{19} V(\phi)&= 3M^2\Mpl^2\kk{1-\f{3+\beta}{6}\ck{1-\cos \mk{\sqrt{2\beta} \f{\phi}{\Mpl} } } } , \\
\phi&=  2\sqrt{\f{2}{\beta}}\Mpl {\rm arctan} (e^{\beta Mt}) , \label{20}\\
\label{21}H&= -M\tanh \mk{\beta Mt}=M\cos \mk{\sqrt{\f{\beta}{2}}\f{\phi}{\Mpl}} , \\
a&\propto \cosh^{-1/\beta} \mk{\beta Mt}=\sin^{1/\beta} \mk{\sqrt{\f{\beta}{2}}\f{\phi}{\Mpl}}. \label{22}
\end{align}
The potential is zero for
\be\label{23}
\phi_c=\f{\Mpl}{\sqrt{2\beta}}\arccos\mk{1-\f{6}{3+\beta}},
\ee
and is negative for $\phi>\phi_c$. This is illustrated in fig.(\ref{fig2}) for $\beta=0.015$
The spectral index, $n_s$, and the tensor-to-scalar ratio, $r$, may be appropriately approximated as \cite{cr2}
\begin{eqnarray}\label{24}
n_s -1 & = -6\epsilon + 2\eta \nonumber \\
r & = 16 \epsilon
\end{eqnarray}
where
\begin{eqnarray}\label{25}
\epsilon & \equiv \f{1}{2}\mk{\f{V^\prime}{V}}^2 \nonumber \\
\label{27}\eta & \equiv \f{V^{\prime\prime}}{V}.
\end{eqnarray}
From \cite{planck2018}, we have :
\begin{align}\label{28}
n_s & = 0.968 \pm 0.006 \nonumber \\
r & < 0.12
\end{align}
So the allowed region for $\beta$ and the scalar field, corresponding to the potentials  \eqref{14} and \eqref{19} are:
\begin{figure}[H]
\centering
\includegraphics[width=7cm]{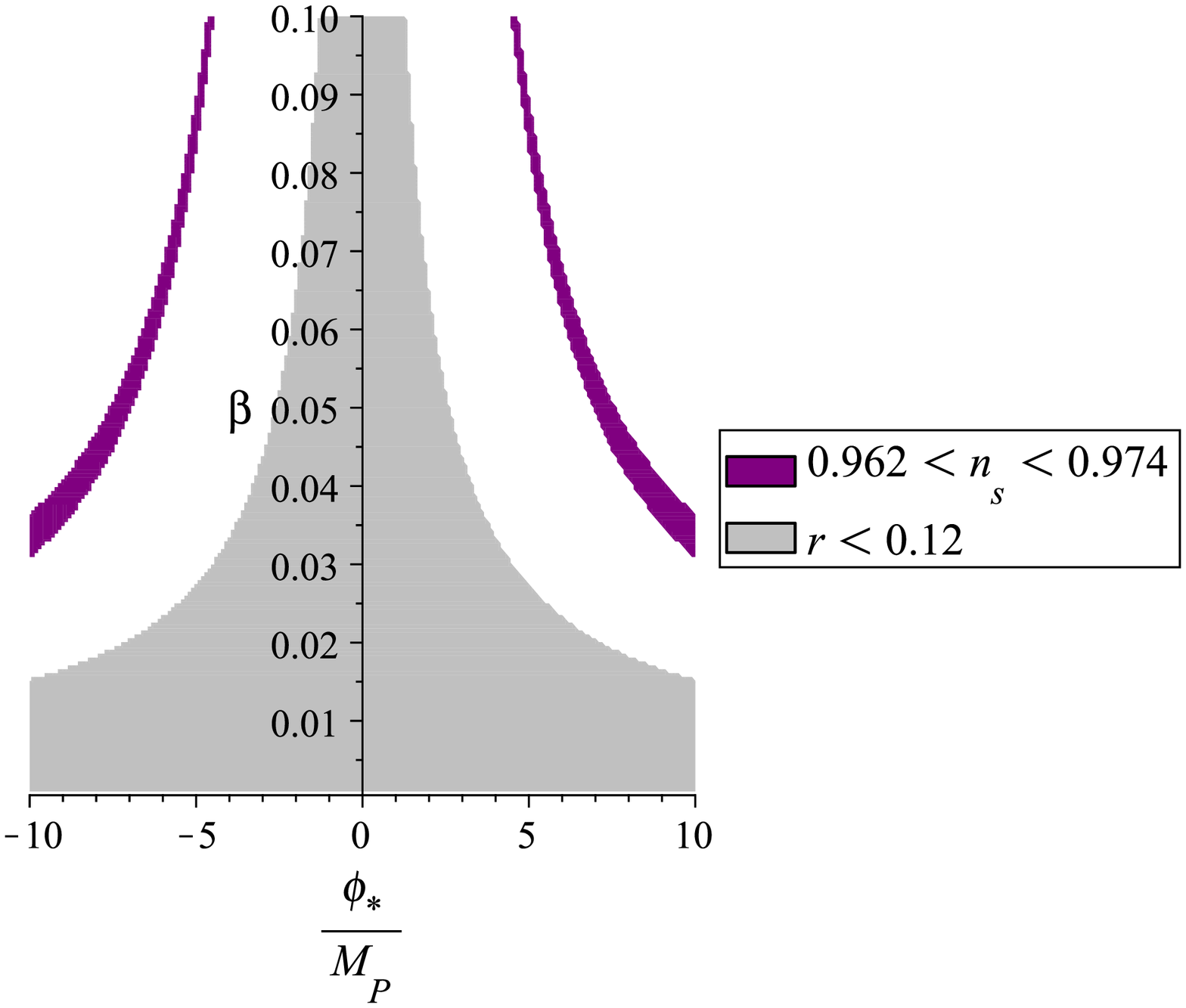}
\caption{The allowed region in the model \eqref{14}.}
\label{fig3}
\end{figure}
and
\begin{figure}[H]
\centering
\includegraphics[width=7cm]{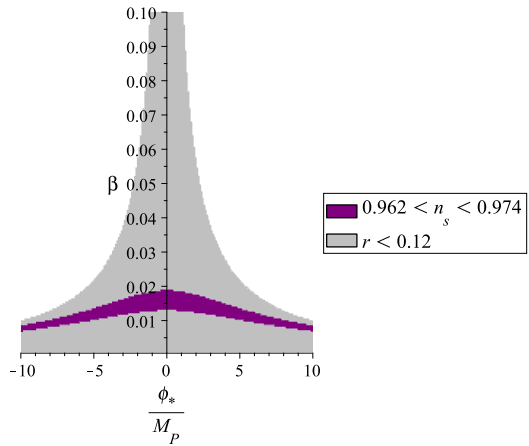}
\caption{The allowed region in the model \eqref{19}.}
\label{fig4}
\end{figure}
As is illustrated in fig.(\ref{fig3}), (\ref{28}) can not be satisfied for the potential (\ref{14}). In the continue we adopt potential (\ref{19}).
\section{End of inflation and reheating temperature }
In this part, by modifying the constant-roll potential, we introduce a mechanism through which the inflation ends and afterwards the Universe warms up. We consider the model specified by the potential (\ref{19}).

In the inflation, the deceleration parameter given by
\begin{equation}\label{r1}
q=-\frac{\ddot{a}a}{\dot{a}^2}=-(1+\frac{\dot{H}}{H^2})=-\left(1-2M_{\rm P}^2\left(\frac{d\ln(H(\phi))}{d\phi}\right)^2\right)
\end{equation}
is positive. For $\beta<0$, and for the potential (\ref{19}), by using $H=M \coth(-\beta Mt)$ and $\dot{H}=-M^2\beta(1-\coth^2(-\beta Mt))$, we find
\begin{equation}\label{r2}
q=-1-\frac{\beta}{\cosh^2(-\beta Mt))}
\end{equation}
So if $q(t_0)<0$ in inflationary era, then we have $q<0$ for $t>t_0$, and the inflation never ends.

For $\beta>0$, we have
\begin{equation}\label{r3}
H=-M\tanh \left( \beta Mt \right) =M\cos x
\end{equation}
where
\begin{equation}\label{r4}
x=\sqrt{\frac{\beta}{2}}\frac{\phi}{M_P}=2\arctan \left( e^{\beta Mt} \right)
\end{equation}
and the deceleration parameter is
\be\label{r5}
q=-1+\beta \tan ^2\left( x \right).
\ee
Here we have $t<0$, because $H$ must be positive. For $t=\to-\infty$, we have $H\to M$ and $x\to 0$. At $t=0$, we have $x=\f{\pi}{2}$ and $H(t=0)=0$ and thereafter a turnaround occurs . In contrast to the slow-roll model, where after a while the slow-roll ceases and an oscillatory phase begins, the constant-roll continues without any oscillating phase. for  $\beta>0$ the inflation ends at $t=\frac{-1}{\beta M}\mathrm{arcsinh}\left( \beta \right)$ where $x=\arctan \left( \frac{1}{\beta} \right)$, and the constant rolls continues toward a turnaround at $t=0$.
		
To end the inflation with a coherent oscillatory phase, we modify the potential as
\be\label{30}
V_{new}(\phi)=f(\phi)V_{old}(\phi)
\ee
where $V_{old}(\phi)$ is the potential of the constant-roll model (\ref{19}). As we aim that the potential be the same as $V_{old}$ in the inflation era, we must have $f(\phi)\simeq 1$ in that epoch. But by evolution of the scalar field $f(\phi)$ deviates from 1 and the inflation ends. $V_{new}$ has a minimum at $\phi_{min}$, such that for $\phi\nsim \phi_{min}$, $V_{new}(\phi)\sim V_{old}(\phi)$ while for $\phi\sim \phi_{min}$  \cite{ch,reh1}:
\be\label{31}
V_{new}(\phi) = \frac{1}{\fact{q}}V_{new}^{(q)}(\phi_{min})(\phi-\phi_{min})^q+\mathcal{O}\mk{(\phi-\phi_{min})^{q+1}}
\ee
where $q$ is the order of the first non-zero derivative of $V_{new}$ at $\phi_{min}$. We assume that $q$ is an even number and for simplicity we have taken $V_{new}(\phi_{min})=0$ . An explicit example will be given in the fourth section. After the inflation, the inflaton begins an oscillation around $\phi_{min}$, and in the background of this rapid oscillating scalar field, relativistic particles are created and reheat the Universe \cite{reh1}.

In order to study the exit from inflation, and to compute the reheating temperature we follow the method used in \cite{mir,sad1,sad2,sad3}. We divide the evolution of the Universe into four parts:
1)$(t_*,t_{end})$, where $t_*$ is when a pivot scale, $\lambda_0$,  exited the horizon and $t_{end}$ denotes the end of inflation. In this period we have a constant-roll inflation.  2) $(t_{end},t_{reh})$, where $t_{reh}$ is the beginning of radiation domination. In this period the inflaton decays to relativistic particles which become eventually in thermal equilibrium at $t_{reh}$ and the radiation dominated era begins.  3) $(t_{end},t_{rec})$, from radiation dominated to the recombination era, and finally :4) $(t_{rec},t_0)$ from the recombination era until the present epoch $t_0$. The number of e-folds since the horizon exit till now can be then written as (\cite{lid})
\begin{align}\label{32}
\mathscr{N} = \ln \mk{\f{a_0}{a_*}} = & \ln \mk{\f{a_0}{a_{rec}}} + \ln \mk{\f{a_{rec}}{a_{reh}}} + \ln \mk{\f{a_{reh}}{a_{end}}} +\ln \mk{\f{a_{end}}{a_*}} \nonumber \\
:= & \mathscr{N}_4 + \mathscr{N}_3 + \mathscr{N}_2 + \mathscr{N}_1
\end{align}

\subsection{The inflationary epoch $(t_*,t_{end})$}
$t_*$ is when a pivot scale, $\lambda_0$, exited the horizon. If we define $k_0=\frac{1}{\lambda_0}$, then by using $a_0k_0=a_*k_*$ we find  $a_* H_* =k_0$ \cite{lid}. We denote by the underlines $"_0"$, and  $"_*"$ the present time and the horizon exit time, respectively. We take $a_0=1$.  In the inflationary era we have
\begin{align}\label{33}
\mathscr{N}_1 &= \int_{t_*}^{t_{end}}{H dt}\nonumber\\
&= \int_{\phi_*}^{\phi_{end}}{\frac{H}{\dot{\phi}} d\phi}
\end{align}
which by using (\ref{7}) becomes
\be\label{34}
\mathscr{N}_1= -\f{1}{2\Mpl^2}\int_{\phi_*}^{\phi_{end}}{\frac{H}{\f{dH}{d\phi}} d\phi}.
\ee
Inserting (\ref{21}) in (\ref{34}) gives
\begin{eqnarray}\label{35}
\mathscr{N}_1 &=& -\f{1}{2}\int_{\tilde{\phi}_*}^{\tilde{\phi}_{end}}{\frac{H}{\f{dH}{d\tilde{\phi}}} d\tilde{\phi}}\nonumber \\
&=& \f{1}{\sqrt{2\beta}}\int_{\tilde{\phi}_*}^{\tilde{\phi}_{end}}\cot \mk{\sqrt{\f{\beta}{2}}\tilde{\phi}} d\tilde{\phi}\nonumber \\
&=& \f{1}{\beta}\ln\mk{\sin\mk{\sqrt{\f{\beta}{2}}\tilde{\phi}}}\Bigl\lvert_{\tilde{\phi}_*}^{\tilde{\phi}_{end}}
\end{eqnarray}
where a tilde denotes a dimensionless parameter obtained through dividing it by $M_P$: $\tilde{\xi}=\frac{\xi}{M_P}$.
So $\mathscr{N}_I$ depends on $\phi_*$, $\phi_{end}, $ and $\beta$. Note that from (\ref{21}) one obtains:
\be \label{36}
\tilde{\phi}_* = \sqrt{\f{2}{\beta}}\arccos \mk{\f{\tilde{H}_*}{\tilde{M}}}
\ee
To compute ${H}_*$, we use the power spectrum of the curvature perturbation \cite{cr1}
\be\label{37}
\Delta_s (k) \equiv \f{k^3}{2\pi^2}\abs{\zeta(k)}^2
= \f{H^2}{8\pi^2 M_P^2 \epsilon_1} \mk{\f{k}{aH}}^3 \f{\pi}{2} \abs{H_\nu^{(1)}(-k\tau)}^2
\ee
in which $\displaystyle\epsilon_1 \equiv -\f{\dot{H}}{H^2}$. From the asymptotic behavior $\displaystyle\lim_{x \rightarrow 0} H^{(1)}_\nu (x)\simeq -\f{i}{\pi}\Gamma(\nu)\mk{\f{x}{2}}^{-\nu}$, we find :
\be\label{38}
\Delta_s (k) =\f{H^2}{8\pi^2 M_P^2 \epsilon_1} \f{2^{2\nu-1}\abs{\Gamma(\nu)}^2}{\pi} \mk{\f{k}{aH}}^{3-2\nu}
\ee
which may be rewritten as:
\be\label{39}
\Delta_s (k) =A_s \mk{\f{k}{aH}}^{n_s -1}
\ee
where:
\begin{align}
& n_s -1 = 3-2\nu \hspace{1cm} \label{40} \\
& A_s = \f{H^2}{8\pi^2 M_P^2 \epsilon_1} \f{2^{2\nu-1}\abs{\Gamma(\nu)}^2}{\pi}\label{41}
\end{align}
At the horizon exit, $k_*=a_*H_*$, hence from $\Delta_s (k_*) =A_s$
\be \label{42}
\Delta_s (k_*) = \f{H_*^2}{8\pi^2 M_P^2 {\epsilon_1}_*} \f{2^{2\nu-1}\abs{\Gamma(\nu)}^2}{\pi}.
\ee
In the constant-roll
\be\label{43}
H=-M\tanh(\beta Mt)
\ee
holds, therefore
\be\label{44}
\epsilon_1 \equiv -\f{\dot{H}}{H^2} = \beta \mk{\f{M^2-H^2}{H^2}}.
\ee
By substituting this in (\ref{42}) we find
\be\label{45}
\Delta_s (k_*) = \f{\tilde{H}_*^4}{8\pi^2 \beta \mk{\tilde{M}^2-\tilde{H}^2}} \f{2^{3-n_s}\abs{\Gamma\mk{\f{4-n_s}{2}}}^2}{\pi},
\ee
where, $\tilde{H}=\f{H}{M_P}$ and $\tilde{M}=\f{M}{M_P}$ and so on. Hence $\tilde{H}_*$ satisfies
\be \label{46}
\tilde{H}_*^4+2C\tilde{H}_*^2-2C\tilde{M}^2=0
\ee
in which
\be\label{47}
C(n_s,\Delta_s(k_*),\beta)=\f{8\pi^3 \beta \Delta_s (k_*)}{2^{3-n_s}\abs{\Gamma\mk{\f{4-n_s}{2}}}^2}
\ee
By noting that $C$ is positive, we find
\be\label{48}
\tilde{H}_* = \mk{-C+\sqrt{C^2+2C\tilde{M}^2}}^{\f{1}{2}}
\ee
If we fix $\Delta_s(k_*)$ and $n_s$ by  astrophysical data. e.g. from \cite{planck2018}  for $k_0=0.05 Mpc^{-1}$ (68\%CL; TT; TE;EE + lowE + lensing))
\begin{align}\label{49}
\ln\mk{10^{10}\Delta_s(k_0)} &= 3.044\pm0.014 \nonumber \\
n_s &= 0.9645\pm 0.0042; \nonumber \\
\end{align}
$\tilde{H}_*$ is determined by specifying only the potential parameter in the inflationary epoch : i.e. $M$ and $\beta$. Note that $\phi_*$ and $\beta$ must be still in the allowed domain, derived from astrophysical data  (see fig.(\ref{fig4}) and related discussions). Equivalently $\tilde{\phi}_*$ may be  expressed as
\be
\tilde{\phi}_*=\sqrt{\f{2}{\beta}}\arccos\mk{\f{\mk{-C+\sqrt{C^2+2C\tilde{M}^2}}^{\f{1}{2}}}{\tilde{M}}}.
\ee
 $\phi_{end}$ depends on $f(\phi)$ in (\ref{30}). We require the inflation ends before the field arrives to $\phi_{c}$, defined in (\ref{23}). Hence generally  $\phi_{end}$ lies between $\phi_*$ and $\phi_{c}$. An appropriate estimation about $\phi_{end}$ is possible only when the form of $f(\phi)$ in (\ref{30})is specified. We leave this topic for section 4, where via an example we elucidate our results.

\subsection{Reheating era $(t_{end},t_{reh})$}
After the inflation the Universe enters the reheating era, $(t_{end},t_{reh})$.  The governing potential is taken as
\be\label{50}
V_{new}(\phi) = \Lambda (\phi-\phi_{min})^q
\ee
where $\Lambda=\frac{1}{\fact{q}}V_{new}^{(q)}(\phi_{min})$.
In this period, the Universe is composed of the inflaton and particles to which the inflaton decays. In the original perturbative approach \cite{reh1}, the scalar field decays through a coherent rapid oscillation around the minimum of its potential, behaving as a matter with the equation of state (EoS) parameter
\begin{equation}\label{51}
w=\frac{q-2}{q+2}.
\end{equation}
But due to collective effects such as the Bose condensation which enhances the decay rate, the primitive perturbative approach is not precise.
Also for large coupling constants and large inflaton amplitude, the
perturbative method fails, and one must consider higher-order Feynman diagrams. In these cases, due to the parametric resonance in the preheating era, a large number of particles are produced. The produced particles evolve from an initial vacuum state in
the background of the oscillating scalar field. A result of this oscillation
is a time-dependent frequency for the produced bosonic fields which
satisfy Hill’s equation \cite{pre2}. From Floquet analysis one obtains
a broad parametric resonance and quick growth of matter in the oscillating inflaton background
\cite{pre2,mu,pod}. The produced particles are initially far from thermal equilibrium but eventually reach thermal equilibrium in the radiation dominated era.

In the perturbative approach, the inflaton gradually decays to relativistic particles, and in the reheating era, the Universe is assumed to be nearly composed of the oscillating inflaton field with the EoS parameter(\ref{51}). But by considering the preheating, this assumption fails [53], and particles created via parametric resonance, must be considered too. In this situation, instead of (\ref{51}),  one may consider an effective EoS parameter $w_{eff.}$
Until the nature of the inflaton and its interactions are identified, a precise analysis of the preheating and deriving an exact form for $w_{eff.}$ is not feasible. Anyway, in the reheating period after the inflation we expect to have $\dot{H}+H^2<0$, which implies $w_{eff.}>-\frac{1}{3}$. If the end of inflation is accompanied by the rapid oscillation $w_{eff.}$ begins by (\ref{51}) , and eventually in the beginning of radiation dominated epoch: $w_{eff.}=\frac{1}{3}$.  The evolution of EoS
between these values has been studied numerically in \cite{pod,mun}. This evolution depends on the effective masses of created particles and their interactions.

In the reheating era, following \cite{pre2}, we estimate the number of e-folds as
\be\label{52}
\mathscr{N}_2 =\ln\mk{\f{a_{reh}}{a_{end}}}=-\f{1}{3\bar{\gamma}}\ln\mk{\f{\rho_{reh}}{\rho_{end}}}
\ee
in which $\bar{\gamma}=\bar{w}_{eff.}+1$ and
\be\label{53}
\bar{w}_{eff.}=\frac{\int_{t_{end}}^{t_{reh}}w_{eff.}dt }{t_{reh}-t_{end}}
\ee
We have also
\be\label{54}
\rho_{end}\simeq 3M_P^2H^2(\phi_{end}),
\ee
and \cite{mu}
\be\label{55}
\rho_{reh}\simeq \f{g_{reh}}{30}\pi^2T^4_{reh}
\ee
where $g_{reh}$ is the number of relativistic degrees of freedom. For Glashow-Weinberg-Salam $g = 106.75$, so $g_{reh}\geq 106.75$.
Inserting (\ref{54}) and (\ref{55}) in (\ref{52}) we obtain
\be\label{56}
\mathscr{N}_2 =-\f{1}{3\bar{\gamma}}\ln\mk{\f{\f{g_{reh}}{30}\pi^2T^4_{reh}}{3M_P^2H^2(\phi_{end})}}
\ee
\subsection{e-folds in $(t_{reh},t_{rec})$ and  $(t_{rec},t_0)$}
In $(t_{reh},t_{rec})$, the Universe is composed of ultrarelativistic particles in thermal equilibrium. The Universe expands adiabatically such that the entropy per comoving volume is conserved, therefore \cite{mir}:
\be\label{57}
\f{a_{rec}}{a_{reh}} = \f{T_{reh}}{T_{rec}}\mk{\f{g_{reh}}{g_{rec}}}^{\f{1}{3}}
\ee
The relativistic degrees of freedom correspond to photons, hence:
\be\label{58}
\mathscr{N}_3 = \ln \mk{\f{T_{reh}}{T_{rec}}\mk{\f{g_{reh}}{2}}^{\f{1}{3}}}
\ee
Finally in $(t_{rec},t_0)$, using the fact that the temperature redshifts as
\be\label{59}
\f{T(z)}{T(z=0)}=\f{a_0}{a}=1+z,
\ee
we simply obtain:
\be\label{60}
\mathscr{N}_4 = \ln \mk{\f{T_{rec}}{T_{CMB}}}
\ee

\subsection{Reheating temperature}
Collecting all the e-folds together
we find
\be\label{61}
\mathscr{N} = \ln \mk{
\mk{\f{\sin\mk{\sqrt{\f{\beta}{2}}\tilde{\phi}_{end}}}{\sin\mk{\sqrt{\f{\beta}{2}}\tilde{\phi}_*}}}^{\f{1}{\beta}}
\mk{{\f{3M_P^2H^2(\phi_{end})}{\f{g_{reh}}{30}\pi^2T^4_{reh}}}}^{3\bar{\gamma}}
\mk{\f{T_{reh}}{T_{CMB}}\mk{\f{g_{reh}}{2}}^{\f{1}{3}}}}
\ee
We remind that $\mathscr{N}$ is the number of e-folds from horizon exit of a pivot scale $k_0$ in inflationary era till now. So by setting $a_0=1$  and by using $a_* H_* =k_0$, we have
\be\label{62}
\mathscr{N} = \ln \mk{\f{1}{a_*}}= \ln \mk{\f{H_*}{k_0}}
\ee
by equating (\ref{61}) and (\ref{62}) we find the reheating temperature as
\be\label{63}
\tilde{T}_{reh}= \mk{
\f{\tilde{k}_0}{\tilde{H}_*}
\mk{\f{\sin\mk{\sqrt{\f{\beta}{2}}\tilde{\phi}_{end}}}{\sin\mk{\sqrt{\f{\beta}{2}}\tilde{\phi}_*}}}^{\f{1}{\beta}}
\mk{{\f{3\tilde{H}^2(\phi_{end})}{\f{g_{reh}}{30}\pi^2}}}^{3\bar{\gamma}}
\mk{\f{1}{\tilde{T}_{CMB}}\mk{\f{g_{reh}}{2}}^{\f{1}{3}}}
}^{\f{1}{12\bar{\gamma}-1}}
\ee
where ${\tilde{H}_*}$ is given by (\ref{48}), and the power spectrum and the spectral index are chosen from (\ref{49}). $g_{reh}\geq 106.75$, and $\phi_{end}$ depends on the form of $f(\phi)$ in (\ref{30}). Note that in the computation of the reheating temperature we have not used explicitly the coupling of the inflaton to particles it decays to.

During this section, for evaluating the perturbations and observables $n_s$ and $r$, the relations (\ref{24}) and (\ref{25}) have been used.
Now let us evaluate $\epsilon$ and $\eta$ for our new model:
\begin{eqnarray}\label{r6}
&&V_{new}^{'}=V_{old}^{'}f+V_{old}f'\nonumber \\
&&\Longrightarrow \,\,\epsilon _{new}=\frac{1}{2}\left( \frac{V_{new}^{'}}{V_{new}} \right) ^2=\epsilon +\left( \sqrt{\epsilon \xi}+\xi \right) \nonumber \\
&&V_{new}^{''}=V_{old}^{''}f+2V_{old}^{'}f'+V_{old}f'\nonumber \\
&&\Longrightarrow \,\,\eta _{new}=\frac{V_{new}^{''}}{V_{new}}=\eta +\left( 4\sqrt{\epsilon \xi}+\sqrt{2\xi} \right)
\end{eqnarray}
where $\xi =\frac{1}{2}\left( \frac{f'}{f} \right) ^2$. If $\xi$ is sufficiently small, it will not have considerable effect and the spectral index and the tensor-to-scalar ratio which are already obtained for the constant-roll model ((\ref{24}) and (\ref{25})) are still acceptable.

\section{A numerical analysis via an example}
\begin{figure}[H]
\centering
\includegraphics[width=8cm]{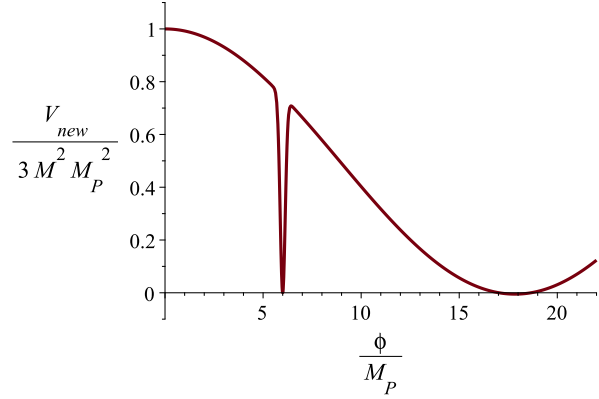}
\caption{Potential (\ref{30}) with parameters (\ref{67}) versus the scalar field.}
\label{fig_V_new}
\end{figure}

\begin{figure}[H]
\centering
\includegraphics[width=7cm]{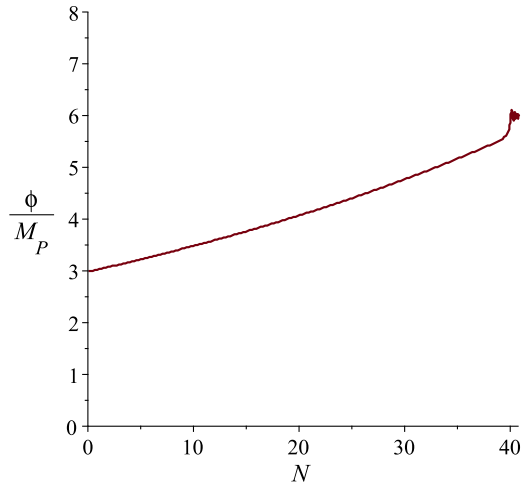}
\caption{Scalar field in terms of {e-folds} $N$, for parameters (\ref{67}) and initial condition (\ref{68}).}
\label{fig_phi}
\end{figure}

\begin{figure}[H]
\centering
\includegraphics[width=7cm]{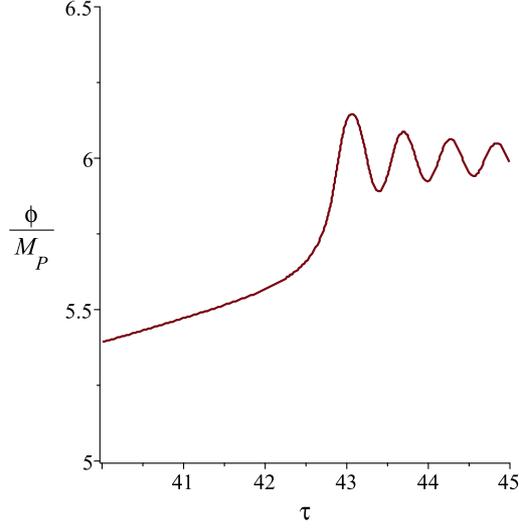}
\caption{Scalar field in terms of $\tau$ in oscillation stage. The chosen parameters and conditions are  \eqref{67} and \eqref{68}.}
\label{fig_phi_t}
\end{figure}

\begin{figure}[H]
\centering
\includegraphics[width=7cm]{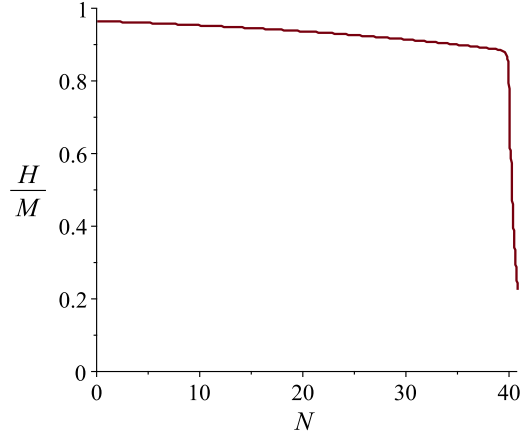}
\caption{Hubble parameter in terms of {e-folds} for parameters (\ref{67}) and initial conditions (\ref{68}).}
\label{fig_H}
\end{figure}

\begin{figure}[H]
\centering
\includegraphics[width=7cm]{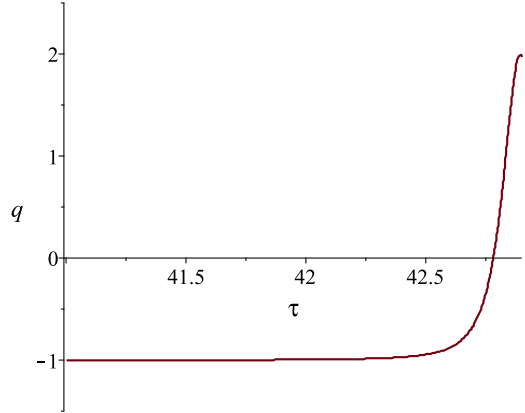}
\caption{Deceleration parameter in terms of dimensionless time $\tau$ for \eqref{67} ‌and (\ref{68}).}
\label{fig_q_t}
\end{figure}

In the following, to be more specific and to elucidate numerically our results, we will make use of a choice for $f(\phi)$, which allows an initial constant-roll evolution, inflation exit and a subsequent oscillation about $\phi\simeq \phi_{\min}$. A good simple example which satisfies the requirements of our model (see eq.(\ref{30}) and its following discussions) is a $f(\phi)$ constructed from an exponential function:
\be\label{64}
f(\phi)=1-e^{-\lambda(\phi-\phi_{min})^q}.
\ee
where $q$ is an even positive number. In this way $f(\phi)$  becomes operative when $\phi\sim \phi_{min}$.
We estimate $\phi_{end}$ as follows: $f(\phi)$ becomes active when
\be \label{65}
\lambda(\phi_{end}-\phi_{min})^q\simeq 1.
\ee
This leads to
\be\label{66}
\phi_{end}\simeq \phi_{min}-\lambda^{-1/q}.
\ee
For the parameters
\be\label{67}
\{ \lambda=30 M_P^{-2}, \hspace{0.5cm} \phi_{min}=6 M_P, \hspace{0.5cm} q=2, \hspace{0.5cm} \beta=0.015 \}
\ee
the potential is plotted in fig.(\ref{fig_V_new}).
Near $\phi_{min}$ we have $V(\phi)=\lambda V_{old}(\phi_{min})(\phi-\phi_{min})^2$ where $V_{old}$ is given by (\ref{19}).  Now, let us solve (\ref{2}) and  (\ref{4}) numerically. We choose the initial conditions
\be\label{68}
\{ \phi_*=3 \, \Mpl, \hspace{0.5cm} \dot{\phi}_*=0.041  \, \Mpl M  \}
\ee
at $t_*$.  Note that, by using (\ref{7}), $\dot{\phi}_*$ becomes fixed by specifying $\phi_*$. By using (\ref{68}), (\ref{35}), (\ref{65}), and for the pivot scale $k_0=0.05 Mpc^{-1}$  we obtain $\mathscr{N}_1\simeq42$. In fig.(\ref{fig_phi}), $\frac{\phi}{M_p}$ is plotted in terms of e-folds $N$ defined by $N=\ln(\f{a(t)}{a_*})$.
As it is shown in fig.(\ref{fig_phi}), the inflaton first ascends during the inflation and then at the end of inflation experiences an oscillatory period around $\phi_{min}=6M_P$. This second stage is plotted separately in fig.(\ref{fig_phi_t}) in terms of dimensionless time $\tau=Mt$. It is in this period that the inflation decays to other particles and reheats the Universe. In plotting fig.(\ref{fig_phi_t}), we have not considered the decay effect via reheating or preheating, and therefore the decrease of the oscillation amplitude is only due to the redshift.
The Hubble parameter in terms of e-folds is depicted in fig.(\ref{fig_H}), showing the constant-roll and the end of inflation after $\mathscr{N}\simeq 42$.
Fig.(\ref{fig_H}) shows a slow change in the Hubble parameter during the inflation, and then its rapid decline $\dot{H}<0$ at the end of inflation. To show the end of inflation, it is convenient to use the deceleration parameter $q$. In terms of the dimensionless time $\tau$, $q$ is depicted in fig.(\ref{fig_q_t}).
This figure shows that the Universe exits the inflation, after the constant-roll, at about $\tau_{end}\simeq 42.78$.

\begin{figure}[H]
\centering\includegraphics[width=7cm]{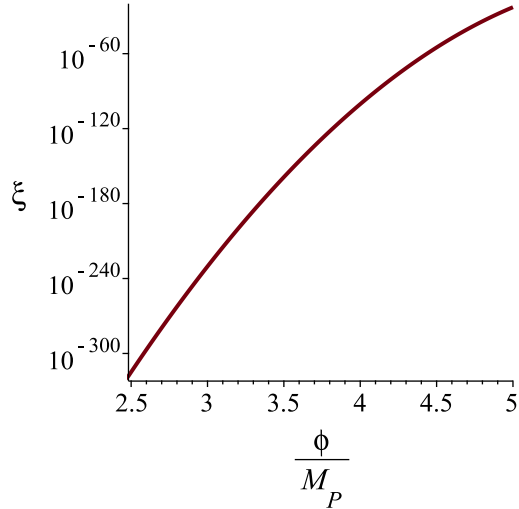}
\caption{$\xi$ in terms of $\f{\phi}{M_P}$ for parameters (\ref{67}).}\label{fig_xi}
\end{figure}

In fig.(\ref{fig_xi}), $\xi$ (defined in (\ref{r6})) is plotted in terms of $\f{\phi}{M_P}$ according to parameters (\ref{67}).
As we can see in fig.(\ref{fig_xi}), $\xi$ is very small and thus would not change (\ref{24}) considerably, therefore we have:
\begin{align*}
\begin{array}{c}
	\epsilon _{new}=\epsilon\\
	\eta _{new}=\eta\\
\end{array}\,\,\Longrightarrow \,\,\begin{array}{c}
	n_{s\left( new \right)}=n_s\\
	r_{new}=r.\\
\end{array}
\end{align*}

\begin{figure}[H]
\centering
\includegraphics[width=10cm]{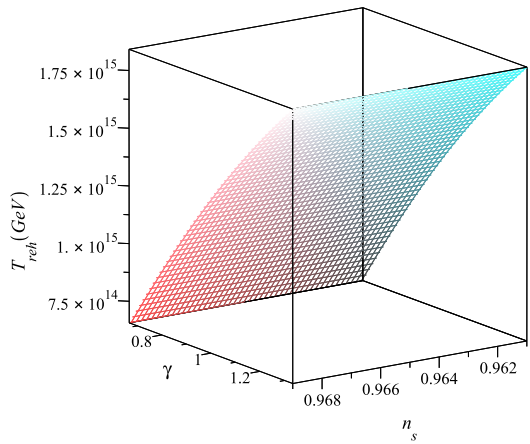}
\caption{Reheating temperature in terms of $\bar{\gamma}$ and $n_s$, for \eqref{67} ‌and \eqref{68}.}
\label{fig_T_reh_3D}
\end{figure}

\begin{figure}[H]
\centering
\includegraphics[width=7cm]{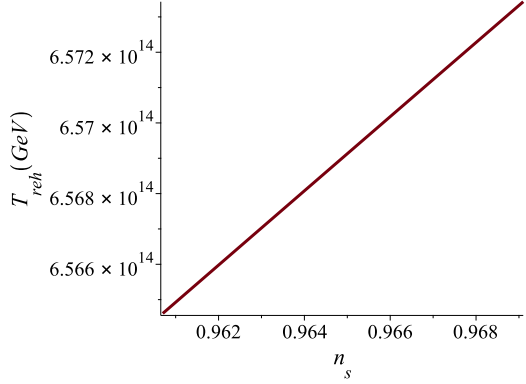}
\caption{Reheating temperature in terms of $n_s$, for $\bar{\gamma}=\f{2}{3}$ for \eqref{67} and \eqref{68}.}
\label{fig_T_reh_2D}
\end{figure}

To get a numerical estimation about the reheating temperature, we depict it in fig.\eqref{fig_T_reh_3D}, in terms of $\bar{\gamma}$ and $n_s$.
We consider (\ref{67}), and the initial condition (\ref{68}), and take  $g_{reh} = 106.75$.
This shows that the temperature does not change significantly in the $n_s$ domain (\ref{49}). We have a greater temperature for a greater value of $\bar{\gamma}$. These can be elucidated separately: in fig.(\ref{fig_T_reh_2D}) the temperature is plotted in terms of $n_s$ for $\bar{\gamma}=\f{2}{3}$, showing that $\frac{\Delta T_{reh}}{T_{reh}}\simeq0.001$. This is in contrast to the slow-roll where the temperature is very sensitive to small changes of $n_s$\cite{mir}.

\begin{figure}[H]
\centering
\includegraphics[width=7cm]{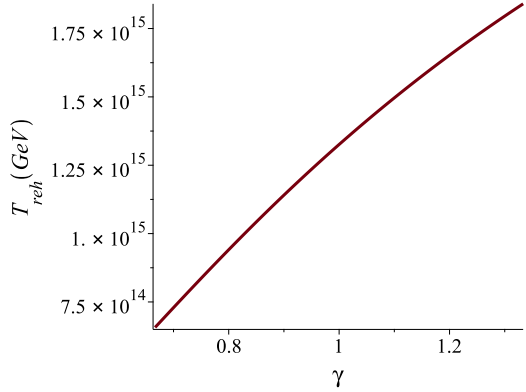}
\caption{Reheating temperature in terms of $\bar{\gamma}$ for $n_s=0.9649$, by considering \eqref{67} and\eqref{68}.}
\label{fig_T_reh_2D_gamma}
\end{figure}

In (\ref{fig_T_reh_2D_gamma}), the temperature  is shown in terms of $\bar{\gamma}$ for $n_s=0.9649$.
As the reheating era occurred symmetry breaking in GUT, we expect that the reheating temperature be smaller than the GUT scale$\sim 10^{16}GeV$.

\begin{figure}[H]
\centering\includegraphics[width=\textwidth]{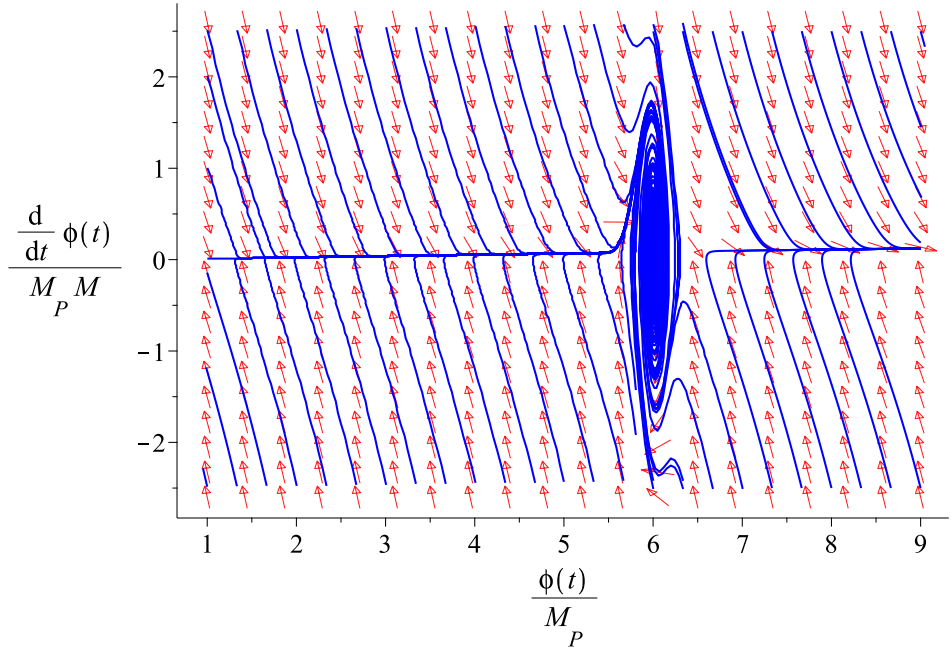}
\caption{ Phase-space diagram for parameters (\ref{67}).}\label{fig_ps}
\end{figure}

At the end, for investigating the dependence of solutions to initial conditions in inflation era, we numerically solved (\ref{2}) and (\ref{4}4) with various initial conditions and obtained the phase-space diagram as depicted in fig.(\ref{fig_ps}) for parameters (71). As we can see in fig.(\ref{fig_ps}), this solution is an attractor for any choice of initial conditions satisfying $0<\phi \left( 0 \right) \le 6M_P$. The final fixed point is $\phi =6M_P, \frac{d\phi}{dt}=0$.

\section{Conclusion}
In the slow-roll model, by the evolution of the inflaton, the slow roll conditions cease and the Universe enters a period of rapid oscillatory phase during which the scalar field decays to ultra-relativistic particles and warms up the Universe.  In the constant-roll model, the constant roll does not cease and we may encounter endless inflation or a turnaround. To study the reheating in this model, one can assume that the scalar field decays to ultra-relativistic particles during the inflation, or chooses other scenarios such as what happens in the slow-roll, i.e. considers an oscillatory phase after the inflation. In the latter case, which is the subject of our study, the potential must be such that the constant roll ends and a reheating era begins. The potential initially corresponds to the constant-roll potential, and then by the evolution of inflaton, its shape changes and gives an end to the constant roll inflation (see (\ref{30})). Also, we made the model so that the scalar field oscillates around the minimum of the potential, reminiscent of the reheating stage after the slow-roll (see (\ref{31})). The implication of this model in the inflationary and reheating eras and the e-folds number was studied explicitly. Based on astrophysical data, we followed the method used in \cite{ref1,mir,sad1,sad2}, to obtain the reheating temperature (see (\ref{63})). In the last section, we used a numerical example to show how the model works. We showed that the model can describe appropriately the end of inflation after a suitable number of e-folds.  We found that the reheating temperature is less sensitive to the spectral index compared with the slow-roll. But it seems that the reheating temperature is large and is only a few orders of magnitudes less than the GUT scale (see fig. (\ref{fig_T_reh_2D_gamma})).
\vspace{2cm}

\section*{Acknowledgment}

V. Anari likes to thank the Iran National Science Foundation (INSF) for the partial financial supports.

\end{document}